\documentclass[twocolumn,showpacs,aps,prl]{revtex4}
\usepackage{amsmath,graphicx}

\newcommand{\ra}{\rangle}
\newcommand{\la}{\langle}
\newcommand{\ain}{\hat{a}_{{\rm in}}}
\newcommand{\aout}{\hat{a}_{{\rm out}}}
\newcommand{\bin}{\hat{b}_{{\rm in}}}
\newcommand{\bout}{\hat{b}_{{\rm out}}}
\newcommand{\hc}{\hat{c}}
\newcommand{\kb}{\kappa_b}
\newcommand{\ka}{\kappa_a}

\newcommand{\hN}{\hat{N}}

\begin{document}

\title{Probabilistic generation of entanglement in optical cavities}

\author{Anders S. S\o rensen}
\altaffiliation{Present address: ITAMP, Harvard-Smithsonian Center for
Astrophysics, Cambridge, Massachusetts 02138} 

\author{Klaus M\o lmer}

\affiliation{QUANTOP,  
Danish National Research Foundation Center for Quantum 
Optics \\ Department of Physics and Astronomy, University of Aarhus, 
DK-8000 Aarhus C, Denmark}

\begin{abstract}
  We propose to produce entanglement by measuring  the
  transmission of an optical cavity.  
  Conditioned on the  detection of a reflected photon, 
  pairs of atoms in the cavity are prepared in 
   maximally entangled states.
  The success probability depends on the cavity parameters, but high
  quality entangled states may be produced with a high
  probability even for cavities of moderate quality.  
\end{abstract}


\maketitle

Quantum
Non-Demolition (QND) measurements on samples of atoms have been
proposed as a means to  entangle atoms in the samples
 \cite{kuzmichTeo,duan}. By QND detection
atoms can be projected into an entangled state 
 by the measurement
and following 
these proposals experiments have produced and verified entanglement of
large atomic samples \cite{kuzmichExp,brian}. Here we propose  to use
a similar QND detection to
entangle pairs of atoms inside an optical cavity. By entangling pairs of atoms 
the present scheme could be used to connect small scale quantum
information processors, e.g. trapped ion crystals, and thereby facilitate the
construction of a large scale quantum computer. 

To our knowledge, all existing proposals for quantum computation
with cavity QED require the strong coupling regime
$g^2/\kappa\gamma\gg 1$, where $g$ is the atom-cavity coupling
constant, $\kappa$ is the cavity decay rate and $\gamma$ is the decay
rate of the atoms. Here we avoid this condition by using a 
probabilistic scheme, which is only successful with a
probability $P_{\rm s}<1$, but where we are certain that we have
produced an
entangled state when we have a click in a
detector. Optical transmission  has already been used in
Ref. \cite{kimble} to 
probe atoms inside a cavity, and here we show that a similar setup can
be used to
entangle two atoms in the cavity.
Probabilistic generation of entanglement has also been
suggested in Refs.\ \cite{cabrillo,lukin,plenio,hong}. Our proposal combines 
advantages of each of these proposals: as in
Refs.\ \cite{cabrillo,plenio,hong} we entangle pairs of atoms (not
samples)  and by using cavities the outgoing
photons are in a well confined mode as in Refs.\ \cite{lukin,plenio,hong},
but unlike Refs.\ \cite{plenio,hong} we do not require the strong coupling
limit $g^2/\kappa\gamma\gg 1$.  

\begin{figure}[b]
  \centering
 \includegraphics[width=7cm]{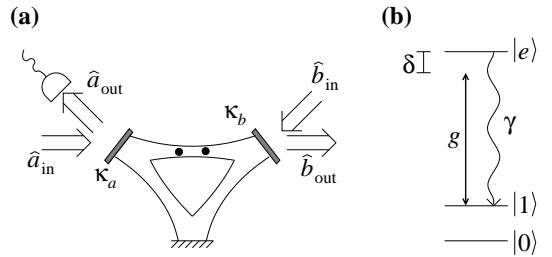}
  \caption[]{Experimental setup and energy levels of the
    atoms. {\bf (a)} Light described by the annihilation operator
    $\ain$ is shined onto a cavity containing two atoms. By
    detecting the reflected light, described by  $\aout$, we
    perform a QND detection of the presence of atoms interacting with
    the field.  {\bf (b)} The atoms
    have two ground states 
    $|0\rangle$ and $|1\rangle$ and an excited state $|e\rangle$ which
    decays to the ground state $|1\rangle$ with a rate $\gamma$. The
    cavity field couples the state $|1\rangle$ to the state
    $|e\rangle$ with 
    a coupling strength $g$ and a detuning $\delta$.}
  \label{fig:setup}
\end{figure}

The experimental setup is sketched in Fig.\ \ref{fig:setup}
{\bf (a)}. We consider a single mode of an optical cavity
 with annihilation operator
$\hc$. In the figure we show a ring cavity, but a standing wave cavity
could also be used. 
The 
field in the cavity can decay through two leaky mirrors with decay
rates $\ka$ and $\kb$. At the mirrors 
we have incoming fields described by $\ain,\bin$
and outgoing fields  described by $\aout,\bout$. 
We assume that light is shined onto the cavity in the mode described
by $\ain$ and  
that the other incoming mode is in the vacuum state. If the
two  cavity mirrors have the same decay rates $\ka=\kb$ and the incident light
is  on resonance with the cavity it is transmitted
through the cavity without any reflection. If, however, 
anything  disturbs the field, the cavity no longer
transmits all the light. By detecting  scattered photons 
we thus register the atomic interaction
with 
the cavity field, and this measurement can be used to entangle 
atoms in the cavity. 

We consider two atoms with two stable ground states $|0\ra$ and
$|1\rangle$ and an excited state $|e\ra$. The cavity field couples the
state 
$|1\ra$ to the excited state 
$|e\ra$ with a coupling strength $g_k$ for the $k$th atom. See
Fig. \ref{fig:setup} {\bf 
  (b)}. The excited state $|e\ra$ 
decays with a rate $\gamma$ and   we
assume a closed transition such
that atoms always end up in 
the state $|1\ra$ after a decay.  Initially each of the atoms is
prepared in the state $\cos(\phi) |0\ra+\sin(\phi)|1\ra$ so that the
combined state of the two atoms is given by
\begin{equation}
  \cos^2(\phi)|00\ra+\sqrt{2} \cos(\phi)\sin(\phi)\frac{
  |01\ra+|10\ra}{\sqrt{2}} +\sin^2(\phi)|11\ra.
\label{eq:inistate}
\end{equation}
Since the state $|0\ra$ does not couple to the cavity
field, the cavity transmits the field with certainty if both atoms are
in this state. If a photon is detected in $\aout$ the $|00\ra$
component of the state 
(\ref{eq:inistate}) is projected out, and if $\phi\ll 1$ we
are left with a  good 
approximation of the maximally entangled state $|\Psi_{{\rm
    EPR}}\ra=(|01\ra+|10\ra)/\sqrt{2}$.

Taking into account realistic imperfections such as
spontaneous emission and imperfect detectors, we calculate the
success probability $P_s$ and the fidelity $F=\la\Psi_{{\rm
    EPR}}|\rho|\Psi_{{\rm 
EPR}}\ra$, where $\rho$ is the density matrix for the atoms. To
calculate these quantities we first calculate the transmission 
properties of the cavity by assuming that the
strength of the field is below saturation, and we then discuss
the effect of photo detection.  

In the rotating frame with respect to the cavity frequency, the
interaction of the atoms with the cavity field is described by the 
Hamiltonian ($\hbar=1$)
\begin{equation}
  \label{eq:ham}
  H=\sum_k g_k |e\ra\la 1|_k \hc+g_k^* \hc^\dagger |1\ra\la e|_k +\delta
  |e\ra\la e|_k,
\end{equation}
where $\delta$ is the detuning of the cavity from the atomic
resonance, and where the sum is over all atoms in the cavity. To
describe the effect of spontaneous emission we 
introduce Lindblad relaxation operators $d_k=\sqrt{\gamma}|1\ra \la
e|_k$ for each of the atoms. 
It is
convenient to adopt different normalization conditions for the
free fields and for the field in the cavity. In the cavity we
normalize $\hc$ to the number of photons whereas the free fields are
normalized to the flux of photons, i.e., the commutation relations are
$[\hc(t),\hc(t)^\dagger]=1$ whereas $[\hat{a}_{{\rm
    in}}(t),\hat{a} _{{\rm in}}(t')^\dagger]=\delta(t-t')$
and similarly for the other free fields. Taking the Fourier transform
of the Heisenberg equations of motion
we
find
\begin{subequations}
  \label{eq:heisenberg}
  \begin{eqnarray}
    \hc(\omega)&=&\frac{\sum_k g_k^* \sigma_{+k}(\omega) 
      +i\sqrt{\ka}\ain(\omega) +i\sqrt{\kb}\bin(\omega) } {\omega +
      i \frac{\kappa}{2}}\label{eq:hc}\\
   \sigma_{+k}(\omega)&=&\frac{\int d\omega' 
       \sigma_{zk}(\omega-\omega')[g_k \hc(\omega')
              -i\sqrt{\gamma} F_k(\omega')]}
            {\sqrt{2\pi}{\left(\omega-\delta+i\frac{\gamma}{2}\right)}}
    \label{eq:be}   \\
  \sigma_{zk}(\omega)&=&\frac{1}{\omega-i\gamma}\bigglb[ -i\gamma
  P_k(\omega)+ 
  \frac{2}{\sqrt{2\pi}}\int d\omega'
  [(g_k^*\hc^\dagger(\omega')+\nonumber \\ &&
  i \sqrt{\gamma} F_k^\dagger(\omega'))\sigma_{+k}(\omega-\omega')- {\rm
    H.C.}]\biggrb], \label{eq:inv} 
  \end{eqnarray}
\end{subequations}
where we have introduced $\sigma_{+k}=|1\ra \la
e|_k$ and $\sigma_{zk}=|1\ra\la 1|_k-|e\ra\la e|_k$ and the projector
onto the $\{|e\ra,|1\ra\}$ subspace $P_k=|1\ra \la 1|_k+|e\ra\la e|_k$.
$F_k$ is the vacuum noise operator associated with the decay of the
$k$th atom, and 
$\kappa=\ka+\kb$. Since the atoms cannot decay out of the
$\{|e\ra,|1\ra\}$ 
subspace the projectors $P_k$ are independent of time and we find
$P_k(\omega)= \sqrt{2\pi}|1\ra\la 1|_k(t=0)\delta(\omega)$, where we
have used that at $t=0$ all atoms are in the ground states. 

To solve the Eqs.\ (\ref{eq:heisenberg})  we assume that the
light intensity is so low  that at most a single photon
is in the cavity at a time.  In this limit the equations of motion
become linear and we can find the 
Fourier transforms of the fields by
solving simple linear equations.   
If we neglect all terms involving more than
a single $\hc$ operator we can omit the integral in Eq.\
(\ref{eq:inv}) when we insert it into Eq.\
(\ref{eq:be}), and by inserting the resulting equation into Eq.\
(\ref{eq:hc}) we obtain
\begin{equation}
  \label{eq:comega}
  \hc(\omega)=\frac{\sqrt{\ka}\ain(\omega)+ \sqrt{\kb}\bin(\omega)+
    i\sqrt{\gamma} \sum_k  \frac{g_k^*P_k(t=0)F_k(\omega)}
    {\gamma/2+i(\delta-\omega)}}{\frac{\kappa}{2}-i\omega+\sum_k
    \frac{|g_k|^2 P_k(t=0)}  {\gamma/2+i(\delta-\omega)}}.
\end{equation}

Using the input/output relations $\aout=\ain-\sqrt{\ka}\hc$ and
$\bout=\bin-\sqrt{\kb}\hc$ we calculate $R(\omega)$ and $T(\omega),$
the reflection and transmission probabilities for incident light 
at a frequency $\omega$. Here we shall only need the
expressions for $R$ and $T$ in the situation where the light is
resonant with the cavity, the cavity mode is resonant with the atoms,
all coupling constants have the same magnitude ($|g_k|^2=g^2$ for all $k$),
and the two mirrors have the same transmittance $\ka=\kb$. In this
situation $R$ and $T$ are
\begin{subequations}
  \label{eq:RT}
  \begin{eqnarray}
    R_{\hN}&=& {\left(\frac{4 \hN g^2/\kappa\gamma}{1+4
          \hN g^2/\kappa\gamma}\right)}^2 \label{eq:R}\\
    T_{\hN}&=& {\left(\frac{1}{1+4
          \hN g^2/\kappa\gamma}\right)}^2 \label{eg:T}.    
  \end{eqnarray}
\end{subequations}
Note that $R$ and $T$ depend on 
$\hN$, the total number of atoms in state $|1\ra$ at $t=0$
($\hN=\sum_k P_k$). 


We assume that the transition between $|1\ra$ and $|e\ra$ is a closed
optical transition so that the 
atoms always end up in the state $|1\ra$ after a decay and
the diagonal density matrix elements
are therefore unaffected by the interaction with the cavity field. The
off-diagonal density matrix elements, however, decay due to
spontaneous emission from the atoms.
From the Hamiltonian (\ref{eq:ham}) and the relaxation 
 operators we
find the equations of motion
\begin{subequations}
  \label{eq:gsderiv}
  \begin{eqnarray}
\frac d{dt} |0\ra\la1|_k&=& {\left(-i
g_k^* \hc^\dagger+\sqrt{\gamma}F_k^\dagger\right)}  |0\ra\la e|_k
\label{eq:01}\\ 
\frac d{dt} |0\ra\la e|_k&=&|0\ra\la1|_k {\left(-i
g_k\hc+\sqrt{\gamma}F_k\right)}.  \label{eq:0e}    
  \end{eqnarray}
\end{subequations}
 To solve the Eqs.\
(\ref{eq:gsderiv}), we integrate Eq.\
(\ref{eq:0e}) with respect to time and substitute it for
$|0\ra\la e|_k$ in Eq.\ (\ref{eq:01}). 
 Assuming that $|0\ra\la1|_k$ does
not change on a time scale $1/\gamma$ and introducing the Fourier
transform of the cavity field we find
\begin{equation}
  \label{eq:01inf}
  |0\ra\la1|_k(t=\infty)=  \ : |0\ra\la1|_k(t=0){\rm e}^{-\int d\omega
  \frac{|g_k|^2
    \hc(\omega)^\dagger\hc(\omega)}{\gamma/2+i(\delta-\omega)}}:,
\end{equation}
where $:\quad :$ denotes normal ordering, and where we have omitted the
noise coming from $F_k$ and $F_k^\dagger$. To evaluate the quality of
the produced entangled state for a pair of atoms we need the coherence
$\xi=|0_1 1_2\ra\la 1_1 0_2| $.  
By using Eq.\ (\ref{eq:01inf}) and the relation between $\hc$
and $\ain$ and by assuming $|g_1|^2=|g_2|^2$ we find 
\begin{equation}
  \label{eq:xi}
  \xi(t=\infty)=\ :\xi(t=0)\exp{\left(-\int d\omega \lambda(\omega)
      \ain(\omega) ^\dagger \ain(\omega) \right)}:
\end{equation}
where $\lambda(\omega)=1-R_1(\omega)-T_1(\omega)$ is the probability
that an incident photon with frequency $\omega$ leads to a spontaneous
emission from an atomic state
with $\hN=1$. Since only resonant light is incident on the cavity
$\lambda=1-R_1-T_1$ with $R_1$ and $T_1$ given by Eq. $(\ref{eq:RT})$
suffices to 
compute the coherence.    


The above expressions completely describe the evolution
of the atoms and fields during the scattering of the light on the
cavity, and we 
now turn to the analysis of the photo detection. 
The theory of photo detection with a finite  
detector efficiency $\eta$ 
can be found in 
Ref. \cite{carmichael}  
($\eta$ also accounts for 
losses between the cavity and the detector, 
e.g., output  mirror loss). We
consider two different types of input fields. First, we consider the
idealized situation where the input field is  a Fock state 
containing a single photon, and later we turn to the more realistic
situation where the input field is in a coherent state.  
With a single incoming photon  and an atomic  state described by
Eq. (\ref{eq:inistate}) the probability to detect the 
photon in the mode $\aout$ is given by
$  P_{{\rm s}}=\eta{\left( p_{1, {\rm i}}R_1+p_{2, {\rm i}}R_2\right)},$
where $p_{1, {\rm i}}=2\sin^2(\phi)\cos^2(\phi)$  and $p_{2, {\rm
    i}}=\sin^4(\phi)$ are 
the initial probabilities to have one and two atoms in state
$|1\ra$, and 
$R_1$ and $R_2$ are the reflection probabilities (\ref{eq:R}). 
 
\begin{figure}[tb]
  \centering
  \includegraphics{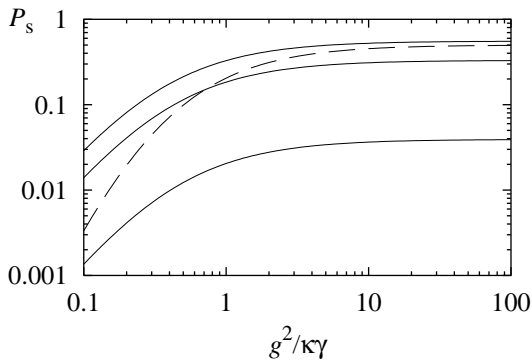}
  \caption{Success probability with a Fock state as input. Full lines
    assume a single detection, the dashed line is for double
    detection. Starting from above the full
    lines have the fidelities $F=0.8$, $0.9$ and $0.99$. Within the
    present model the fidelity of the dashed line is
    unity. The probabilities assume a detector efficiency of
    unity.  With non-ideal detectors the full (dashed) lines should be
    multiplied by $\eta$ ($\eta^2$).}
  \label{fig:fock}
\end{figure}

As a measure of the quality of the produced entanglement we use the fidelity
$F=\la \Psi_{{\rm EPR}}|\rho|\Psi_{{\rm EPR}}\ra=p_{1,{\rm c}}/2+{\rm
  Re}(\xi)$,  
where $p_{1,{\rm c}}$ is the probability to have one atom in state $|1\ra$
conditioned on the detection of a photon. With a Fock state as input
${\rm Re}(\xi)=p_{1,{\rm c}}/2$ because there can not have been spontaneous
emission when the photon is detected by the detector, 
and conditioned on the detection
of a photon the fidelity of the entangled state is 
 $ F=(p_{1, {\rm i}}R_1)/(
p_{1, {\rm i}} R_1+p_{2, {\rm i}}R_2).$
In Fig.\ \ref{fig:fock} we show the success probability $P_{\rm s}$ 
 as a function of $g^2/\kappa\gamma$ for $\eta=1$ with
fixed
values of the fidelity $F=0.8$, $0.9$, and $0.99$. 
With non-ideal detector efficiency the probability
should be multiplied by $\eta$.

If a high fidelity of the entangled state is required ($F\approx 1$) the
success probability becomes very low, and  it is advantageous
to use a double detection scheme  in which we prepare the
atoms in the initial state (\ref{eq:inistate}) with $\phi=\pi/4$
($p_{1, {\rm i}}=1/2$).
If we detect a photon we interchange the 
states $|0\ra$ and $|1\ra$ in both atoms and we probe the transmission 
of the cavity
once more. If another photon is detected we have excluded both the
$|00\ra$ and $|11\ra$ components of Eq.\ (\ref{eq:inistate})
 and we are left with an entangled state with a fidelity
$F=1$. The success probability for the double detection $P_{\rm
  s}=\eta^2R_1^2/2$ is 
also plotted in Fig.\ \ref{fig:fock} (with $\eta=1$). 

It should be noted that imperfections such as absorption in
the mirrors or a mismatch between the incident field and the cavity
mode, may cause reflection of photons which will be mistaken for 
successful generation of the entangled state. If the imperfections cause  
a fraction $f$ of the incident photons to be reflected, the state 
conditioned on the two-photon detection will have its fidelity reduced
by the ratio between the two-count probability in the desired state
and the total two-count probability,
$(R_1(1-f)+f)^2/((R_1(1-f)+f)^2+f(R_2(1-f)+f))$. 
For small values of  $f$ the fidelity is also reduced by a small
amount, i.e., for $g^2/\kappa\gamma=1$ we get $F=0.98$  (0.85) for
$f=1\%$  (10\%).

The photon number states are hard to
produce experimentally, and we shall now investigate the more
realistic situation where the incoming light is in a coherent
state. With coherent light there is a probability to have more than
one photon in the pulse, and thus a probability that the atoms have
spontaneously emitted a photon when we detect a photon in the
detector. We assume that the atoms are initially prepared in the state
(\ref{eq:inistate}) and we then continuously monitor the reflection from
the cavity with a continuous source of coherent light in the incoming
mode. If we
have a click in the detector the experiment is successful and we
block the light to avoid 
spontaneous emission.
 If we have not registered a click in the detector
after a certain  mean photon number $n_{{\rm max}}$ have been
shined onto the cavity, the protocol is unsuccessful and we have to
restart the experiment. 


If the first click is observed after a mean photon number
$n$, 
the probability to be in the subspace with one atom in state $|1\ra$
is 
\begin{equation}
  \label{eq:pcond}
 p_{1,{\rm c}}=\frac{p_{1, {\rm i}} R_1{\rm e}^{-\eta R_1
   n}}{p_{1, {\rm i}}R_1{\rm e}^{-\eta R_1 n}+ p_{2, {\rm
     i}}R_2{\rm e}^{-\eta 
   R_2 n}}.
\end{equation}
According to Eq.\ (\ref{eq:xi}) the atomic coherence is reduced by the 
factor ${\rm e}^{-\lambda n}$ after the interaction with a coherent
state with a mean photon number $n$, and we find
$   F_{\rm c}=p_{1, {\rm c}}/2+{\rm Re}(\xi)={p_{1, {\rm
       c}}}(1+{\rm e}^{-\lambda n})/2.$
By averaging this expression with the probability distribution for the
first click
$ dP_{\rm f}/dn=\eta p_{1, {\rm i}}R_1{\rm e}^{-\eta R_1 n}+ \eta
  p_{2, {\rm i}}R_2{\rm e}^{-\eta R_2 n}  $
we find the average fidelity 
\begin{equation}
  \label{eq:Fc}
  F=p_{1, {\rm i}}\frac{{\left(1-{\rm e}^{-\eta R_1 n_{\rm
            max}}\right)}+\frac{\eta R_1}{\eta R_1+\lambda} {\left(1-{\rm
          e}^{-(\eta 
          R_1+\lambda) n_{\rm max}}\right)}} {2P_{\rm s}},
\end{equation}
where the success probability $P_{\rm s}$ is given by
\begin{equation}
  \label{eq:Pc}
  P_{\rm s}= p_{1, {\rm i}}{\left(1-{\rm e}^{-\eta R_1 n_{\rm
          max}}\right)}+ p_{2, {\rm i}}{\left(1-{\rm e}^{-\eta R_2
        n_{\rm max}}\right)}.
\end{equation}
 
\begin{figure}[t]
  \centering
  \includegraphics{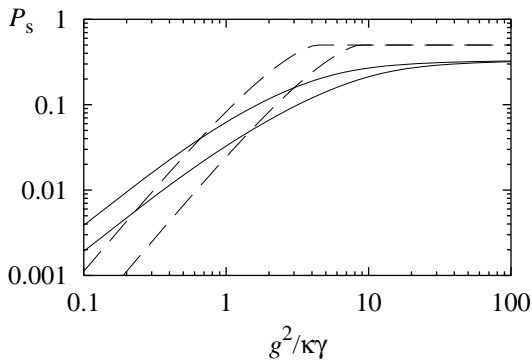}  
  \caption{Success probability for $F\geq 0.9$ with coherent light as
    an input. Full 
    lines assume a single detection, and dashed lines are for double
    detection. The upper (lower) curves are for a detector efficiency
    $\eta=1$ ($\eta=0.5$).}
  \label{fig:coherent}
\end{figure}

For a given cavity the fidelity and  success probability depend on the
preparation angle $\phi$ and the maximum mean 
photon number  after which the preparation is
restarted $n_{\rm max}$. We have numerically optimized the success
probability with a 
fixed fidelity. The optimal probability for $F=0.9$ is shown with full
curves in Fig.\
\ref{fig:coherent}. The upper (lower) full curve is for a
detector efficiency $\eta=1$ ($\eta=0.5$). The preparation angle
$\phi$ varies between 0.2 and 0.4 over the range of the figure, and
$n_{\rm max}$ is on the order of unity for $g^2/\kappa\gamma\sim 1$
(to be precise: all the curves in  the figure have $n_{\rm max}<2$ for
$g^2/\kappa\gamma<2 $). 

Like for Fock states, it is an advantage to use a
double detection scheme for high fidelities and/or good
cavities. Again we prepare the atoms in the initial state
(\ref{eq:inistate}) with $\phi=\pi/4$. If we have a click after
a  mean number of photons  $n_1\leq n_{1,{\rm max}}$ have been sent in, we
interchange the states $|0\ra$ and $|1\ra$ and wait for a second
click. The scheme is successful if a second photon is detected
after a mean number of photons  $n_2\leq n_{2,{\rm max}}$ have  been shined
onto the cavity. 
We find that the
optimal strategy is to take the scheme to be successful if
$n_1+n_2\leq n_{{\rm max}}$. With this strategy we find the fidelity 
\begin{equation}
  \label{eq:Fcfc}
  F=\frac{1}{2}+\frac{\eta^2R_1^2{\left(1-(1+(\eta R_1+\lambda)n_{\rm
          max})  {\rm e}^{-(\eta R_1+
          \lambda)n_{\rm max}} \right)}}{2(\eta R_1+\lambda)^2 P_{\rm s}},
\end{equation}
where the success probability $P_{\rm s}$ is given by
\begin{equation}
  \label{eq:Pcfc}
  P_s=\frac 1 2 {\left(1-{\rm e}^{-\eta R_1
          n_{\rm max}}(1+ \eta R_1 n_{\rm
          max})\right)}
\end{equation}
In Fig.\ \ref{fig:coherent} the dashed curves show the success probability
$P_{\rm s}$ for the double detection when we require $F\geq 0.9$ (for
the flat part of the curves $F$ is actually larger than 0.9).  

Strong
coupling is usually attempted in standing wave cavities but recently a ring
cavity with a finesse of $1.8\cdot 10^5$   has been reported in
\cite{hemmerich}, and we estimate this cavity to be in the regime 
with $g^2/\kappa\gamma\sim 1$ if the beam waist is reduced to $30 \mu$m.
For the ring cavity our scheme has the additional advantage
that it only depends on the magnitude of the coupling
constant $|g|^2$ and not  on the phase, and hence, the atoms  only need to be 
confined within the cavity waist and not within a wavelength of the field.
In the ring cavity we must, however, take the mode traveling in the
opposite direction into account. The precise effect of a photon
leaving the cavity from this mode depends on the separation
between the atoms and 
unless we restrict this separation to a half-integer number of 
wavelengths of the radiation,
such an event will be as harmful as
spontaneous emission. Treating photons leaving from the
counter propagating mode on equal footing with spontaneous emission
events, they have the effect of replacing $g^2/\kappa\gamma$ by
$g^2/(\kappa\gamma+4\tilde{g}^2\kappa/\tilde{\kappa})$ in the above
expressions, where $\tilde{g}$ and $\tilde{\kappa}$ are the coupling 
constant and decay rate for the counter propagating mode. 
If $g=\tilde{g}$ and $\kappa=\tilde{\kappa}$ this limits the
effective $g^2/\kappa\gamma$ to values less than 1/4. 
Inserting an optical diode, however, may significantly increase
$\tilde{\kappa}$ and thereby eliminate the role of the
counter propagating mode.

In conclusion we have proposed a realistic scheme for 
entanglement of atoms inside an optical cavity. The proposed scheme does
not require the cavity to be in the strong coupling regime
$g^2/\kappa\gamma \gg 1$ and high quality entangled pairs can be produced
for cavities with $g^2/\kappa\gamma\sim 1$. 
 The entanglement protocol
has a finite success probability, and hence one can with certainty
produce the entangled state by simply trying sufficiently many times.
This implies that we can carry out quantum gates
by using the entanglement as a channel for teleportation of the 
qubit contents which we assume to be stored  in other quantum mechanical 
degrees of freedom, e.g., another ion in an ion trap.


 \end{document}